\documentstyle[12pt]{article}
\catcode`@=11
\newcount\@tempcntc
\def\@citex[#1]#2{\if@filesw\immediate\write\@auxout{\string\citation{#2}}\fi
  \@tempcnta\z@\@tempcntb\m@ne\def\@citea{}\@cite{\@for\@citeb:=#2\do
    {\@ifundefined
       {b@\@citeb}{\@citeo\@tempcntb\m@ne\@citea\def\@citea{,}{\bf ?}\@warning
       {Citation `\@citeb' on page \thepage \space undefined}}%
    {\setbox\z@\hbox{\global\@tempcntc0\csname b@\@citeb\endcsname\relax}%
     \ifnum\@tempcntc=\z@ \@citeo\@tempcntb\m@ne
       \@citea\def\@citea{,}\hbox{\csname b@\@citeb\endcsname}%
     \else
      \advance\@tempcntb\@ne
      \ifnum\@tempcntb=\@tempcntc
      \else\advance\@tempcntb\m@ne\@citeo
      \@tempcnta\@tempcntc\@tempcntb\@tempcntc\fi\fi}}\@citeo}{#1}}
\def\@citeo{\ifnum\@tempcnta>\@tempcntb\else\@citea\def\@citea{,}%
  \ifnum\@tempcnta=\@tempcntb\the\@tempcnta\else
   {\advance\@tempcnta\@ne\ifnum\@tempcnta=\@tempcntb \else \def\@citea{--}\fi
    \advance\@tempcnta\m@ne\the\@tempcnta\@citea\the\@tempcntb}\fi\fi}
\catcode`@=12

\def\barr{\begin{array}}
\def\earr{\end{array}}
\def\beq{\begin{equation}}
\def\eeq{\end{equation}}
\def\bea{\begin{eqnarray}}
\def\eea{\end{eqnarray}}

\def\bmath{\begin{displaymath}}
\def\emath{\end{displaymath}}
\def\bq{\begin{quote}}
\def\eq{\end{quote}}
\def\Re{\mbox{Re}}
\def\Im{\mbox{Im}}

\def\cL{{\cal L}}

\def\cT{{\cal T}}
\def\<{\langle}
\def\>{\rangle}

\def\apprle{\hspace{-0.1cm}\stackrel{\displaystyle <}{\sim}}

\def\slash#1{\setbox0=\hbox{$#1$}#1\hskip-\wd0\hbox to\wd0{\hss\sl/\/\hss}}

\def\npb#1{{\em Nucl.\ Phys.\ }{\bf B#1}}
\def\plb#1{{\em Phys.\ Lett.\ }{\bf A#1}}
\def\plb#1{{\em Phys.\ Lett.\ }{\bf B#1}}
\def\prl#1{{\em Phys.\ Rev.\ Lett.\ }{\bf #1}}

\def\prd#1{{\em Phys.\ Rev.\ }{\bf D#1}}

\begin{document}

\begin{flushright}
RAL-94-119\\[-0.2cm]
CPP-94-34\\[-0.2cm]
DOE-ER-40757-057 \\[-0.2cm]
November 1994
\end{flushright}

\begin{center}
{\bf{\LARGE Signatures of Higgs-Triplet Representations}}\\[0.3cm]
{\bf{\LARGE at TeV-{\boldmath $e^+e^-$} Colliders}}\\[2.5cm]
{\large Kingman Cheung}$^a${\large ,
Roger J.N.~Phillips}$^b${\large , and Apostolos Pilaftsis}$^b$\\[0.4cm]
{\it $^a$University of Texas at Austin, Center for Particle Physics,}\\
{\it Austin, TX 78712, USA}\\[0.3cm]
{\it $^b$Rutherford Appleton Laboratory, Chilton, Didcot, Oxon, OX11 0QX, UK}
\end{center}
\vskip2cm
\centerline{\bf ABSTRACT}

We investigate the potential of future TeV linear $e^+e^-$ colliders to
observe singly-charged Higgs bosons ($H^\pm$) via the coupling
$H^\pm W^\mp Z$, which would signal the existence of exotic
Higgs representations. In the context of a Higgs-triplet model
compatible with the electroweak oblique parameters, we estimate the cross
section for producing charged Higgs-triplet bosons that couple
predominantly to $W$ and $Z$ bosons in 0.5--2 TeV-$e^+e^-$
colliders. The principal backgrounds are evaluated and the viability
of the signal is discussed and illustrated.

\newpage

The physics potential of a TeV-$e^+e^-$ collider, such as the proposed
next linear collider (NLC) planned to operate with a centre-of-mass
energy of 500~GeV~\cite{NLC}, includes the exploration of
the parameter space of theories beyond the minimal Standard
Model (SM).  Theories beyond the SM usually predict the existence of
charged Higgs bosons ($H^\pm$)~\cite{SK}.
In particular, if charged  Higgs bosons are produced via
a {\em sizeable}  $H^\pm W^\mp Z$ coupling, this alone can reveal
the origin of the charged Higgs boson, {\em i.e.}, as a member of a
Higgs-triplet realization beyond the SM.
In extensions of the SM with Higgs doublets and singlets, the coupling
$H^\pm W^\mp Z$ vanishes at tree level and can
only be generated at one-loop level~\cite{GKW,PHI}.
The reason for the vanishing of the $H^\pm W^\mp Z$ coupling in the
Lagrangian is rather technical and depends upon the hypercharge
($Y$) and weak-isospin assignments of the Higgs representations
introduced in the model. In fact, the $W^\pm$ and $Z$ bosons couple
through the covariant derivative to the charged Higgs and would-be
Goldstone ($G^\pm$) bosons, and a tree-level coupling to the singlet Higgs
fields ($Y=0$) is therefore prohibited. On the other hand,
in models with complex Higgs doublets $\Phi_i$ ($Y=1$), the vertex
$H^\pm W^\mp Z$ is proportional to $T_- \< \Phi_i\>$ and hence
vanishes [$T_\pm=\frac{1}{2}(\sigma_1 \pm \sigma_2)$, with the Pauli matrices
denoted by $\sigma_{1,2,3}$], whereas $G^\pm W^\mp Z \propto
T_+\<\Phi_i\> $ is non-zero as should be the case in a renormalizable
extension of the SM.
In multi-Higgs doublet models, the resulting strength of the loop-induced
$H^+W^-Z$-coupling turns out to be rather small of the order of $10^{-2}$
relative to the SM vertex $HW^+W^-$.  A  large $H^\pm W^\mp Z$ coupling
is therefore an  indicator of exotic triplet or higher
Higgs representations beyond the SM;  searches for experimental
signatures of this coupling will offer  unique tests for
the presence of such exotic representations.

In the context of theories containing $Y=2$ Higgs-triplet fields,
our aim is to show that TeV-$e^+e^-$ colliders are capable of
differentiating whether the charged Higgs bosons belong to
a triplet or doublet representation after taking into account
the SM background.  Such a distinction is harder to achieve at hadron
colliders; searches there for doublet charged Higgs bosons have
been discussed~\cite{BHP}. Complex triplet representations also predict
doubly charged Higgs bosons ({\em i.e.}~$H^{++}$); we shall not
address their signatures here, but
refer the reader to Ref.~\cite{VD} for $H^{++}$ signals at hadron
colliders and Ref.~\cite{e-e-} for
$H^{--}$ production at $e^-e^-$ linear colliders.

In models with Higgs triplets, one has to face difficulties coming
from large contributions to the electroweak parameters $S$, $T$, and
$U$~\cite{PT} (generalized to $V$, $W$, and $X$~\cite{CPB}).
Especially, compatibility with the Veltman parameter $\rho$
($\propto T$)~\cite{MV} and the absence of large flavour-changing
neutral currents suggest that the neutral component of the left-handed
triplet should possess an unnaturally small vacuum expectation value
of the order of eV. An interesting scenario that avoids this problem
was considered by Galison~\cite{PG}, and Georgi and Machacek~\cite{GM}.
They introduced more than one triplet field into
the model and imposed an $SU(2)$ custodial symmetry on the vacuum
expectation values and hypercharges of the Higgs multiplets
to ensure $\rho=1$
at tree level. This scenario was further analyzed by Chanowitz
and Golden~\cite{CG}, who examined stability conditions of the $SU(2)$
custodial symmetry in the Higgs potential under higher order quantum
corrections. To be more precise, the model under discussion consists of
the usual SM $Y=1$ complex doublet $\Phi$, plus one real $Y=0$ and one
complex $Y=2$ triplet given by
\beq
\Delta\ =\ \left( \barr{ccc}
\delta^0 & \chi^+ & \delta^{++} \\
\delta^- & \chi^0 & \delta^+ \\
\delta^{--} & \chi^- & \delta^{0\ast} \earr \right).
\eeq 
Among the various Higgs fields that the model predicts,
there exist charged Higgs-triplet bosons $H^\pm$ (sometimes denoted
as $H^\pm_5$), which have no-tree level couplings to fermions.
In addition to $H^\pm$, the model also contains charged Higgs-doublet bosons
$H'^\pm$ (also called $H^\pm_3$) that do not couple to gauge bosons
in the Born approximation. Specifically, after diagonalizing the charged
Higgs-boson matrix by assuming that the $SU(2)$ custodial symmetry
is preserved, they are identified as
\beq
H^\pm = \sqrt{\frac 12 }\, (\delta^\pm -\chi^\pm ), \qquad
H'^\pm = c_H \sqrt{\frac 12}\, (\delta^\pm +\chi^\pm ) - s_H \phi^\pm,
\eeq 
where $\phi^+$ is the charged-field component of the Higgs doublet
$\Phi$ and $s_H=\sqrt{1-c^2_H}$ is the sine of a doublet-triplet mixing angle
defined as
\beq
s_H\ =\ \sqrt{\frac{8v_T^2}{v_D^2+8v_T^2}}, \label{sH}
\eeq 
with $v_D/\sqrt{2}=\<\phi^0\>$ and $v_T=\<\delta^0 \>
=\<\chi^0 \>$. The SM vacuum expectation value is then related
to $v_D$ and $v_T$ via $v^2=v_D^2+8v_T^2$.
The corresponding vertex $H^+W^-Z$ is then given by~\cite{GHKD}
\beq
\cL_{int} \ =\ -g_w \frac{s_H}{c_w} M_W H^+W^{-\mu}Z_\mu\ +\ H.c.,
\label{int}
\eeq 
where $g_w$ is the usual $SU(2)_L$ electroweak coupling constant,
$c_w^2=1-s^2_w=M^2_W/M^2_Z$ and $s_w^2=\sin^2\theta_w$ is the sine-square
of the Weinberg angle.   Due to electromagnetic gauge invariance,
the coupling $H^\pm W^\mp \gamma$ is absent at tree level. As emphasized
earlier, we are interested in a large $H^\pm W^\mp Z$ coupling that will
unavoidably signify the triplet nature of the charged Higgs bosons $H^\pm$.
This can only be the case if $s_H \sim 1$ or equivalently
$v_T \sim v_D$, which is considered to be a natural scenario.
In the limit of our interest ($s_H\to 1$), the only interactions of
$H^\pm$ with other fields that survive
in the Lagrangian are those between the so-called fiveplet members
($H_5^{--}, H^-, H^0_5, H^+, H^{++}_5$ in the notation
of~\cite{GRW}), {\em i.e.},\ the couplings $H^0_5H^-W^+$, $H^+H^{--}_5W^+$,
$H^+H^- Z$, $H^- H^+ \gamma$, and the one given by Eq.~(\ref{int}).
There is also a tri-Higgs vertex $H^+ H'^- H^0_3$, which depends
crucially on the details of the Higgs potential. Thus, for some
specific choice of parameters, $H^0_3$ can be heavier than $H^+$ and
$H^+ \not\to H'^+ H^0_3$.
An exhaustive list of the Feynman rules containing all the Higgs particles
involved in this model can be found in~Ref.~\cite{GRW}. Furthermore,
as an effect of the $SU(2)$ custodial symmetry, all fiveplet members
are degenerate in mass and so the only dominant decay mode is
$H^+\to W^+Z$.  The partial width of this decay channel
is given by
\bea
\label{GH}
\Gamma (H^+\to W^+ Z) & = & \frac{\alpha_w}{16} \left(\frac{s_H^2}{1}\right)
\; M_H \lambda^{1/2}(M^2_H/M^2_W,1/c^2_w,1)\,
\Big[ 1+ x_W^2 + x_Z^2 \nonumber\\
&&-2 x_W -2 x_Z +10 x_W x_Z \Big] \;,
\eea
with $\alpha_w=g^2_w/4\pi$, $\lambda(x,y,z)=(x-y-z)^2-4yz$, $x_W=M_W^2/M_H^2$,
and $x_Z=M_Z^2/M_H^2$.
Of course, if such a scenario were embedded in a grand unified theory
(GUT), one would have to cope with the known gauge-hierarchy problem
or problems related to the existence of a unification point at the
GUT scale $M_X$. Solutions to these problems may be achieved by
considering a supersymmetric GUT version that contains our
low-energy model~\cite{GRW}. One may therefore
expect that additional supersymmetric scalars
will be present in the theory and give rise to new decay modes for
a very heavy charged Higgs boson with mass of order 1 TeV.
For our present illustrations, however, it is reasonable to
consider a scenario in which $B(H^+ \to W^+ Z) \simeq 1$
for charged Higgs masses $M_H\apprle 600$ GeV and $s_H\simeq 1$.

There are two preferred channels for hunting the charged-triplet Higgs
at TeV-$e^+e^-$ colliders: (i) $e^+e^- \to Z^*\to W^-H^+$
and (ii) $e^+e^-\to W^{+*} Z^* e^- \bar \nu_e \to H^+ e^- \bar \nu_e$
(illustrated in Fig.~1), both of which depend on the $W^\pm Z H^\mp$ vertex.

\vspace{0.7cm}
\noindent
{\em (i) The Bjorken-type process $e^+e^- \to Z^*\to W^- H^+$}\\[0.5cm]
The Feynman diagram is depicted in Fig.~1(i).
The total cross section for
$e^+e^- \to Z^* \to W^{\pm *}H ^\mp \to f \bar f' H^\mp$, where $W^{\pm *}$
denotes an off-shell $W$ boson and $f\bar f'$ is any fermion pair from the
$W$ decay, is given by
\beq
\label{6}
\sigma_{tot}(s)\ =\ \frac{\alpha_w^3 N_c s^2_H}{192 c^4_w s}\
\frac{(1+v_e^2)M^2_W}{(s-M^2_Z)^2\ +\ M^2_Z\Gamma^2_Z}\ I(s,M^2_H),
\eeq
where $v_e=1-4s^2_w$, the colour factor $N_C=3\ (1)$
for quarks (leptons), and
\bea
I(s,M^2_H) &=& \Re J(2M_H\sqrt{s},s+M^2_H-M^2_W-i\Gamma_W M_W,s+M^2_H)
\nonumber\\
&&+\ \frac{M_W}{\Gamma_W} \Im J(2M_H\sqrt{s},s+M^2_H-M^2_W-i\Gamma_W M_W,
s+M^2_H ).
\eea
The function $J(\alpha,\beta,\gamma)$ is defined as
\bea
J(\alpha,\beta,\gamma) &=& \int\limits_{\gamma}^{\alpha} dx
\frac{\sqrt{x^2-\alpha^2}}{x-\beta}\
\ =\ -\sqrt{\gamma^2-\alpha^2}-i\pi\sqrt{\beta^2-\alpha^2}
+(\beta+\sqrt{\beta^2-\alpha^2} ) \ln\alpha \nonumber\\
&& -\beta \ln (\gamma +\sqrt{\gamma^2-\alpha^2}) -
\sqrt{\beta^2-\alpha^2}\ln\left( \frac{\beta\gamma - \alpha^2 -
\sqrt{\beta^2-\alpha^2}\sqrt{\gamma^2-\alpha^2}}{\gamma -\beta} \right).\quad
\eea
In case of complex arguments, the function $J(\alpha,\beta,\gamma)$
should be continued analytically.  In Fig.~2, we have
plotted the total cross section (summing $H^+$ and $H^-$ channels) as a
function of the charged Higgs mass at center-of-mass energies
$\sqrt{s}=0.5, 1, 1.5$ and 2 TeV.  We also summed over all the fermion
pairs $f\bar f'$.
Since $H^\pm$ decays into $W$ and $Z$ bosons, the process of interest
is $e^+e^- \to f\bar f' W^\pm Z$, in which
the vector bosons may be identified via their leptonic decays into
electrons and muons.  Obviously, the irreducible background is the SM
production of $e^+e^- \to W^+W^- Z$.
The leptonic branching fractions are:
\beq
B( W^- \to e^-, \mu^- + X_\nu) \ \simeq\ 0.26,\qquad
B( Z \to e^- e^+, \mu^- \mu^+ ) \ \simeq\ 0.067, \label{BR}
\eeq
where we have included the modes $W^-\to \tau^- \bar \nu_\tau \to
e^-/\mu^- \bar \nu_{e/\mu} \nu_\tau \bar \nu_\tau$, so
the quantity $X_\nu$ denotes either one or three neutrinos.
{}From these branching fractions and the cross sections of Fig.~2,
we see that pure leptonic signals from charged Higgs
production via this process are very small, and decrease
as c.m.s.~energy $\sqrt s$ increases as indicated by the $1/s$ factor
in Eq.~(\ref{6}).
We shall therefore henceforth focus our attention mainly on the
more promising $WZ$ fusion process, but we will take account of small
contributions  from the Bjorken process at $\sqrt s = 0.5$ TeV,
where they are not  negligible.

\newpage
\noindent
{\em (ii) The fusion process $e^+e^-\to W^{+*} Z^* e^- \bar \nu_e \to
H^+ e^- \bar \nu_e$} \\[0.5cm]
This reaction depicted in Fig.~1(ii) offers larger signals than the
previous process. The total cross section can be written
\beq
\sigma_{tot}(s)\ =\ \frac{1}{1024\pi^4 s^2}
\int \frac{ds_2dt_1dt_2ds_1}{\sqrt{-\Delta_4(s,s_1,s_2,t_1,t_2)}}
\overline{|\<\cT \>|}^2, \label{sig2}
\eeq
where the squared transition element averaged over the spins of the
initial states is
\beq
\overline{|\<\cT \>|}^2 \ =\ \frac{g^2_wM^2_Ws_H^2g_L^{(w)2}}{c^2_w
(t_1-M^2_Z)^2(t_2-M^2_W)^2}
\Big[ g_L^{(z)2} s(s-s_1-s_2+M^2_H) + g_R^{(z)2}(s-s_1+t_2)(s-s_2+t_1) \Big],
\label{T2}
\eeq
and the coupling constants $g_L^{(w)}$, $g_{L,R}^{(z)}$ are given by
\bea
g^{(i)}_{L,R} &=& g_V^{(i)}\ \mp \ g_A^{(i)} ,
\qquad i=w,z \nonumber\\
g^{(w)}_V&=&- g^{(w)}_A\ =\ \frac{g_w}{2\sqrt{2}},\qquad
g^{(z)}_V\ =\ -\frac{g_w}{4c_w}\left(1 - 4s^2_w \right),
\qquad g^{(z)}_A\ =\ \frac{g_w}{4c_w}. \label{gs}
\eea
Here, the superscripts $(w)$ and $(z)$ refer to the production vertices
of a virtual $W$ and $Z$ boson, respectively. Furthermore,
$s$, $s_1$, $s_2$, $t_1$, and $t_2$ in Eq.~(\ref{T2}) are the
usual Mandelstam variables defined as follows:
\bea
s = (k_{e^-}+k_{e^+})^2, \quad
t_1 = (p_{e^-}-k_{e^-})^2, \quad
t_2 = (p_{\nu}-k_{e^+})^2, \nonumber\\
s_1 = (p_{e^-}+p_H)^2, \quad
s_2 = (p_{\nu}+p_H)^2. \label{Mand}
\eea
The phase-space limits of the Mandelstam variables
listed in Eq.~(\ref{Mand}) as well as the definition
of the kinematic function $\Delta_4$ in Eq.~(\ref{sig2})
can be found in Ref.~\cite{BK}.
In Fig.~3, we show the computed total
cross section as a function of the charged Higgs boson mass $M_H$
at $\sqrt{s}=0.5,1,1.5$, and 2 TeV.
Unlike the Bjorken-type process this channel has a  cross
section increasing with $\sqrt{s}$.

The signal of interest is $e^+e^- \to e^\mp \nu H^\pm \to e^\mp
\nu W^\pm Z$ with leptonic
decays; we therefore concentrate on the channel
\beq
e^+e^- \to e\nu WZ \to e\nu (\ell\nu) (\ell' \bar \ell'),
\eeq
where $\ell,\ell'$ denote $e$ or $\mu$.
It is understood that decays $W\to \tau\nu \to \ell\nu\nu\nu$
are always included, since they are practically impossible to be
distinguished experimentally from the direct leptons in
$W\to\ell\nu$, but decays $Z\to\tau\tau\to\ell\ell
\nu\nu\nu\nu$ can be excluded because the dilepton invariant mass is
generally much less than $M_Z$.  The $WZ$-fusion process then has net
branching fraction 0.017, that multiplies the cross section of Fig.~3
to give the cross section in this leptonic channel.

The main characteristics of the $WZ$-fusion four-lepton signal are:
three hard central leptons from $W$ and $Z$ decay; one scattered beam
$e^{\pm}$; two of the leptons reconstruct the $Z$ boson;
and the two undetectable neutrinos
give large missing $p_T$.   Note that the Bjorken process also contributes
in this channel, albeit at a low level, and must be added to the final
signal.

We must now discuss the main SM backgrounds in the above channel,
together with possible kinematic cuts to reduce them with minimal
loss of signal.  These backgrounds are:\\[0.3cm]
(a) $e^+e^-\to W^+ W^- Z$, with one $W$ boson decaying to $e$ or $\mu$
and the other only to $e$.  This background cannot easily be removed
and must be calculated in detail, though the cross section decreases as
$\sqrt{s}$ increases.  It is part of the annihilation channel
$e^+e^- \to W^+ W^- Z^* \to W^+ W^- \ell \bar \ell \;
(\ell=e,\mu)$.  Actually, it can also be viewed as part of
$e^+e^-\to W^- Z W^{+*} \to W^- Z \ell^+ \nu$, or
$e^+e^-\to W^+ Z W^{-*} \to W^+ Z \ell^-\bar \nu$.
To avoid double counting we include it in the
$e^+e^-\to W^- Z W^{+*} \to W^- Z \ell^+ \nu$ calculation.
\\[0.3cm]
(b) $e^+e^-\to e^+e^- W^+ W^-$ with leptonic $W$ decays.
This background refers only to the scattering channel contribution;
the annihilation channel is already included in process (a).
The total cross section of this production is very large, of order 2 pb
at $\sqrt{s}=1.5$~TeV due to the double photon-exchange diagrams.
Fortunately, this  huge cross section can be substantially reduced
by requiring both the scattered beam electron and positron to be away
from the beam direction ({\it e.g.} requiring $|\cos\theta_e| <0.98$),
and by constraining the invariant mass of one lepton pair to be
around the $Z$ mass while the invariant mass of the other pair is
larger than $M_Z+10$~GeV. After all these requirements this
background remains non-negligible, so
we include it in our analysis.
\\[0.3cm]
(c) $e^+e^-\to e^\pm W^\mp Z \nu$, followed by the subsequent decays
$W^\pm\to e^\pm,\mu^\pm X_\nu$ and $Z\to ee,\mu\mu$.
The Feynman graphs of this SM reaction may be found in Fig.~4 of
Ref.~\cite{BCKP}.
This process refers only to the scattering channel, while the annihilation
channel is already included in process (a).
The total cross section of this process is also very large due to
the photon-exchange diagrams.  The cross section can be reduced by
excluding electrons close to the beam,  but the reduction is less
than in $e^+e^-\to e^+e^- W^+W^-$ and
it remains a major background to our signal.
Since this process~\cite{BCKP,HKM} has very similar features to our
signal a more sophisticated investigation of kinematic variables
is needed.  Thanks to the  difference  that there are no
resonance graphs with a heavy charged Higgs boson in this background,
we can exploit the invariant mass of the three charged leptons that decay
from the $WZ$.  While the background should be smooth in this distribution,
the signal should be concentrated in a limited range depending on
the charged Higgs-boson mass. \\[0.3cm]
(d) $e^+e^-\to ZZ$ with leptonic $Z$ decays. The case where both $Z$
decay directly to $e$ or $\mu$ pairs can be suppressed by requiring that
only one pair has invariant mass near $M_Z$, and by requiring a large
missing transverse momentum $\slash{p}_T$.
There remains a contribution where the second $Z$
decays via $Z\to \tau^+\tau^- \to ee,\ e\mu + X_\nu$; this is important
only at $\sqrt s =$0.5 TeV,  where $\sigma_{b} \simeq 0.2$ fb, and
can be removed completely by requiring that the second pair of leptons
have invariant mass greater than $M_Z$.\\[0.3cm]
(e) $e^+e^-\to Ze^+e^-$ with $Z$ decaying directly to $ee$ or $\mu\mu$.
This background refers to the scattering channel (process (d)
already includes the major annihilation channel).
The total production cross section is of order 1 pb at
$\sqrt{s}=0.5-2$~TeV \cite{e-e-}, including the $Z$ decay branching ratio.
It is reduced to the level of 1~fb by cutting out leptons at small beam
angles, and can be finally eliminated by a missing transverse momentum
cut.\\[0.3cm]
(f) $e^+e^-\to ZZZ^*,\ ZZ\gamma^*$. These annihilation processes are
of higher order than process~(d) and therefore generally smaller.
If the final $Z$ bosons and/or the off-shell photon go
to $\ell^+\ell^-$, $\nu\bar \nu$ and $e^+e^-$, respectively, they contribute
to the same final states as (g) and (h) below.\\[0.3cm]
(g) $e^+e^-\to ZZe^+e^-$ scattering, with  one $Z$ decaying invisibly.
This can fake signal events but is at least an order of magnitude smaller
than (c) and the small-angle cut on both the scattered $e^+$ and $e^-$
reduce it to a negligible level. For example, at 1.5~TeV this background
is only of order  $10^{-3}$ fb.\\[0.3cm]
(h) $e^+e^-\to ZZ\nu\nu$ scattering, with one $Z$ boson decaying via
$\tau$ leptons into $e^+e^-\nu\nu\nu\nu$ or  $e\mu\nu\nu\nu\nu$.
This background can be removed by requiring the invariant mass of
$e^+e^-/e\mu$ to be larger than $M_Z$. \\[0.3cm]
Thus the only major backgrounds are (a), (b), and (c).

Our strategies to select the signal and minimize these backgrounds
are as follows.
\begin{itemize}
\item We select events with exactly four charged leptons in the final state
(no hadrons), at least one of which must be $e^{\pm}$, and impose the
following lepton acceptance cuts:
\begin{equation}
p_T^\ell > 10\; \mbox{GeV}\qquad \mbox{and} \qquad
|\cos\theta_\ell| < 0.98 \;
\end{equation}
where $\theta_\ell$ is the angle between the lepton and the beam direction.

\item Since two of the four charged leptons should come from a $Z$ boson, we
require one pair of oppositely charged leptons of the same flavour to
reconstruct the $Z$ mass in the range
\begin{equation}
M_Z - 10\; \mbox{GeV} < M(\ell^+ \ell^-) < M_Z + 10\; \mbox{GeV}\;.
\label{cut1}
\end{equation}

\item For the other pair of leptons (which should come from $W$ decay and
a scattered $e^-/e^+$),  we require them to have  opposite charges, one of
them to be $e^{\pm}$, and their invariant mass $M(e\ell)$ to be above
the $Z$ range:
\begin{equation}
M(e\ell) > M_Z + 10 \; \mbox{GeV}\;.
\end{equation}

\item We impose a missing transverse momentum cut
\beq
\slash{p}_T > 30~\mbox{GeV}.
\eeq

\item We attempt to form the invariant mass of the two leptons,
which reconstruct the $Z$ boson,  plus the lepton from the $W$ decay.
For $e\mu Z$ final states it is uniquely
determined that $M(\mu Z)$ is the correct combination.
But for $eeZ$ final states (half of our signal) the choice is
ambiguous; here we choose the minimum of the two invariant masses
$M(eZ)$, denoted by $M(\ell\ell\ell-min)$.  In the case of the signal,
$M(\ell\ell\ell-min)$ turns out to have a distribution very similar
to the ``correct" invariant mass $M(\mu Z)$  in the $e\mu Z$ channel;
both have the same sharp upper limit
\beq
M^2 < {1\over 2}[M_H^2+M_Z^2-M_W^2+\lambda^{1/2}(M_H^2,M_Z^2,M_W^2)]
    < M_H^2
\eeq
The lower limit on $M(\mu Z)$ is found by reversing the sign of
$\lambda^{1/2}$ above.
This variable is intended to distinguish further between signal
and background.
\end{itemize}

A possible additional strategy would be to select only $e\mu Z$ in the final
state.  This would trivially remove some of the backgrounds and would
remove the need for the $M(\ell\ell\ell-min)$ variable.  However, the
signal would then be halved and the major backgrounds would remain, reduced
by no more than the same factor 2.  We do not choose this option here.

We have computed the triplet-Higgs signal and the main backgrounds
with the above acceptance criteria, using Monte Carlo methods.  The
signal calculations are based on spinor trace techniques; the
$H^{\pm}\to W^\pm Z \to \ell^\pm \nu\ell' \bar \ell'$
decay trace is analogous to
the production trace, with appropriate crossings; the effects of
$W\to\tau\nu \to \ell\nu\nu\nu$ cascade decays are included by the
methods of Ref.\cite{trace}.  The background calculations are based on
helicity amplitude techniques, extending the codes originally
developed in Ref.\cite{BCKP}.
We have restricted ourselves to masses $M_H > M_W + M_Z$ , for
which on-shell $H^\pm \to W^\pm Z$ decays are possible.
Our integrated cross section results are exhibited in Table~1.
Several comments should be made.\\
(i) The signals do not rise monotonically with energy,
unlike the uncut cross sections in Fig.~3. This is mostly because
of the angular cut on the scattered beam electron or positron,
that removes a larger fraction of electrons at higher energy.
The signal would increase if
this cut were relaxed, and ideally one might consider different
cuts for different energies; however, the background would increase
even more (and there are also practical difficulties in detecting
at small angles in linear colliders), so we have not pursued
this option.\\
(ii) The other cuts do not cost more at higher energy. The cut
on the two non-$Z$ lepton invariant mass is in fact the most costly
at the lowest energy, $\sqrt s=0.5$ TeV, where it typically halves
the signal; this is understandable, because the scattered beam
electron is less energetic at lower $s$.\\
(iii) The Bjorken process contributes significantly at the lowest
energy only, giving $20\%$ ($60\%$) of our signal for $M_H=175$ GeV
($350$ GeV) there.\\
(iv) $W\to\tau\nu\to \ell\nu\nu\nu$ decays give between $4\%$ and
$12\%$ of our signal, losing a larger fraction to the cuts, especially
at lower energies.

Are such signals detectable above the backgrounds?
Assuming annual luminosity 50~fb$^{-1}$ at each energy,
and a net lepton identification efficiency of 60$\%$ or more per
event, we see the possibilities at $\sqrt s= 0.5$ TeV are rather
limited; however, charged Higgs-triplet bosons with masses up to
about 400~GeV might eventually be detectable for large values of
the mixing angle $s_H$, at the higher energies.  For example, at
$\sqrt s =1.5$ TeV with $M_H=175-400$ GeV and $s_H\sim 1$, there
would be about 8-12 signal events on top of 9 background  events per
year,  giving a somewhat significant excess in one year.

If the presence of a signal can be detected as an excess of events
over the expected background,  its origin as a $WZ$ resonance can be
confirmed and the Higgs mass extracted by a study of the trilepton
invariant mass distributions.  In Fig.~4, we illustrate the
$M(\ell\ell\ell-min)$ distribution at $\sqrt s=1.5$ GeV, for
triplet Higgs masses 200, 300, 400, 500 GeV; the case of $M(\mu Z)$,
that can only be defined in $e\mu Z$ channels, is rather similar.
We see that the signal and background have quite different distributions.
In the case $M_H=200$ GeV, the narrow signal peak between
$M(\ell\ell\ell - min)=100$ and $M(\ell\ell\ell - min)=180$ GeV
is particularly striking, compared to the broad
background continuum.   For higher Higgs masses, the signal peak
is broader but nevertheless has a sharp upper limit and changes the
net distribution shape in a very significant way.
For estimating the significance of the signal, we should compare
only with the background events directly under the signal peak; this
improves the numerical significance of our signal. For example,
for $M_{H}=200$ GeV at $\sqrt{s}=1.5$ TeV with 50~fb$^{-1}$ luminosity,
we should compare 12 signal events in the Higgs peak with about
4 background events under this peak (see Fig.~4), rather than
the total of 9 background events altogether.

We now briefly discuss the effects of initial state radiation
(bremsstrahlung and beamstrahlung), that are not included in our
analysis above. Both bremsstrahlung and beamstrahlung reduce
the center-of-mass energy $\sqrt{s}$ to an effective
center-of-mass energy $\sqrt{\hat s}$, while beamstrahlung at $e^+e^-$
colliders also increases the effective luminosities.  The effect of
beamstrahlung on the effective luminosities at various $e^+e^-$ collider
designs can be found in Ref.~\cite{chen}; the increase in luminosities
varies from a factor of 1.3 to 3.3 and is favourable to our signal.
The reduction in the effective center-of-mass energy does not have such
an adverse effect on our signal as one might at first suppose; although
the uncut signal cross section decreases with $\sqrt{\hat s}$, this is
compensated by the effect of the cuts, at least at the higher energies
(see Table 1).  Furthermore, although bremsstrahlung is inevitable,
standard, and independent of the collider designs, the beamstrahlung can
always be minimized by designs, {\it e.g.} by using a ribbon-shaped beam.
Thus initial state radiation has only a marginal effect in our analysis,
and can even increase the signal.

Finally, we remark briefly on  the possibility
of using the hadronic decays of $WZ\to (jj)(jj)$, where $j$ denotes a
hadronic jet, to identify the charged Higgs boson.  The advantages of
the hadronic mode are the much larger branching fraction and  the full
reconstruction of the charged Higgs boson.  The increase in branching
ratio is more than a factor of 25. However, the same is true for the
backgrounds, and might be even worse due to additional QCD backgrounds; also
it is impractical to distinguish event-by-event between the $W$ and $Z$
bosons using the hadronic mode, since they give very similar dijet
invariant masses.   Therefore, we have to face much larger backgrounds
from $e^+e^- \to e^+e^- W^+W^-$ and $e^+e^- ZZ$.  There are also
complications due to the other charged Higgs bosons
$H'^\pm$, which do not decay into $WZ$ but mainly into quark jets.
However, if we can reconstruct the hadronic $W$ and $Z$ bosons fairly
cleanly, it should still be possible to distinguish between  $H'^\pm$
and $H^\pm$ production. This possibility might be worth exploring
in the future.

In conclusion, we have investigated the feasibility of using the $WZ$
fusion process, $e^+e^- \to W^* Z e\nu \to H^\pm e^\mp \nu$, to
detect an exotic charged Higgs boson.  If the coupling
$W^\pm Z H^\mp$ is large enough, {\it e.g.}, the case when $H^\pm$ belongs
to a Higgs-triplet and the mixing angle $s_H$ is close to 1, the production
of $H^\pm$ by $WZ$ fusion, followed by $H^\pm \to W^\pm Z \to \ell^\pm \nu
\ell' \bar \ell'$, give a sizeable number of signal events above a few
SM background events.  In addition, we have shown that the invariant
mass distribution $M(\ell\ell\ell-min)$ is a good indicator to test for
the existence of such a singly-charged triplet-Higgs boson.

\vspace{1cm}
\noindent
{\bf Acknowledgements.} Helpful discussions with Tao Han and David Miller
are gratefully acknowledged.
K.C. was supported in part by a DOE grant number
DOE-ER-40757.

\newpage

\newpage

\centerline{\bf\Large Figure and Table Captions }
\vspace{-0.2cm}
\newcounter{fig}
\begin{list}{\bf\rm Fig. \arabic{fig}: }{\usecounter{fig}
\labelwidth1.6cm \leftmargin2.5cm \labelsep0.4cm \itemsep0ex plus0.2ex }

\item Feynman diagrams responsible for producing the
      singly-charged Higgs-triplet
      boson: (i) $e^+e^-\to Z^*\to W^- H^+$ and
      (ii) $e^+e^-\to W^{+*} Z e^- \bar \nu_e \to H^+e^- \bar\nu_e$.

\item Production cross section of the charged Higgs-triplet boson
      via the Bjorken-type process
      $e^+e^-\to Z^*\to W^{\mp *} H^\pm \to f\bar f' H^\pm$
      for different  c.m.s.~energies:
      $\sqrt{s}= 500$ GeV (solid line), 1 TeV (dashed line),
      1.5 TeV (dash-dotted line), and 2 TeV (dotted line).
      It is summed over all possible $f\bar f'$ pairs.

\item Production cross section of the charged Higgs-triplet boson
      via the $WZ$ fusion process $e^+e^-\to H^\pm e^\mp \nu$ for
      c.m.s.~energies
      $\sqrt{s}= 500$~GeV (solid line), 1 TeV (dashed line),
      1.5 TeV (dash-dotted line) and 2 TeV (dotted line).

\item Histogram estimates indicating the excess of the leptonic signal
      from $WZ\to H^{\pm}$ fusion (shaded area) above the background  at
      $\sqrt{s}=1.5$~TeV, as a function of the leptonic
      invariant mass $M(lll-min)$ defined in the text,
      with charged Higgs-boson masses:
      (a)~$M_H=200$~GeV, (b)~$M_H=300$~GeV,
      (c)~$M_H=400$~GeV, and (d)~$M_H=500$~GeV.

\end{list}
\newcounter{tab}
\begin{list}{\bf\rm Tab. \arabic{tab}: }{\usecounter{tab}
\labelwidth1.6cm \leftmargin2.5cm \labelsep0.35cm \itemsep0ex plus0.2ex }

\item Production cross section $\sigma/s^2_H$ (in fb) of charged Higgs bosons
      in the channel $e^+e^-\to H^{\pm *} e^\mp \nu \to W^\pm Z e^\mp \nu;\
      W \to e\nu,\mu\nu, e\nu\nu\nu,\mu\nu\nu\nu,$ and $Z\to ee,\mu\mu$
      as a function of $M_H$ after the kinematic cuts discussed in the text.
      At the end of the table, we also present results for the background
      processes (a), (b), and (c) in fb.

\end{list}

\newpage

\centerline{\bf\Large Table 1}
\vspace{0.5cm}
\begin{tabular*}{13.7218cm}{|r||r|r|r|r|}
\hline
$M_H\ [$GeV$]$
& $\sqrt{s}=0.5$~TeV & $\sqrt{s}=1$~TeV & $\sqrt{s}=1.5$~TeV
& $\sqrt{s}=2$~TeV \\
\hline\hline
{\em \underline{Signal}} & & & & \\
175 & 0.19 & 0.43 & 0.41 & 0.32 \\
200 & 0.15 & 0.40 & 0.39 & 0.31 \\
250 & 0.10 & 0.35 & 0.36 & 0.29 \\
300 & 0.05 & 0.29 & 0.34 & 0.28 \\
350 & 0.02 & 0.24 & 0.31 & 0.28 \\
400 &      & 0.21 & 0.27 & 0.25 \\
500 &      & 0.12 & 0.22 & 0.22 \\
600 &      & 0.06 & 0.17 & 0.18 \\
700 &      & 0.02 & 0.12 & 0.15 \\
800 &      &      & 0.08 & 0.12 \\
\hline\hline
{\em \underline{Background}} & & & & \\
(a)   & 0.04 & 0.08 & 0.07 & 0.06 \\
(b)   & 0.03 & 0.05 & 0.05 & 0.05 \\
(c)   & 0.01 & 0.08 & 0.17 & 0.23 \\
\hline
Total & 0.08 & 0.21 & 0.29 & 0.34 \\
\hline
\end{tabular*}

\end{document}